\documentclass[12pt]{article}
\usepackage{amssymb}
\begin{document}
\mbox{}\\[2cm]
\begin{center}
{\Large \bf Quasiperiodic packings of decagonal two-shell clusters}\\[5mm]
{\it Nicolae Cotfas}\\[5mm]
Faculty of Physics, University of Bucharest, Romania\\
E-mail: ncotfas@yahoo.com\\
http:\verb#//fpcm5.fizica.unibuc.ro/~ncotfas#\\[2cm]
\end{center}
{\bf Abstract.} We present some mathematical results concerning the strip 
projection method and a computer program for generating quasiperiodic packings 
of decagonal two shell-clusters.\\[5mm]

{\large \bf 1. The two-shell decagonal cluster  $\mathcal{C}$}\\[5mm]  
The relation
\[ 
C5:\mathbb{R}^2 \longrightarrow \mathbb{R}^2:  
\left(
\begin{array}{c}
\alpha \\[2mm]
\beta 
\end{array}\right) \mapsto  
C5\left(
\begin{array}{c}
\alpha \\[2mm]
\beta 
\end{array}\right)
\!=\!\left(
\begin{array}{lr}
\cos \frac{2\pi }{5} & -\sin \frac{2\pi }{5}\\[2mm] 
\sin \frac{2\pi }{5} & \cos \frac{2\pi }{5} 
\end{array} \right)
\left(
\begin{array}{c}
\alpha \\[2mm]
\beta 
\end{array}\right) \!=\!
\left( 
\begin{array}{c}
\frac{\sqrt{5}-1}{4}\alpha - \frac{\sqrt{5+\sqrt{5}}}{2\sqrt{2}}\beta \\[2mm]
\frac{\sqrt{5+\sqrt{5}}}{2\sqrt{2}}\alpha + \frac{\sqrt{5}-1}{4}\beta 
\end{array} \right) 
\]
defines the usual two-dimensional representation of the
{\it rotation group} $C_5$.\\
By starting from the given vector
\[ BASIS(:,1)=\left( \begin{array}{c}
BASIS(1,1)\\[2mm]
BASIS(2,1)
\end{array}\right)  \]
we generate the vectors
\[\begin{array}{l}
BASIS(:,2)=C5 \ BASIS(:,1) \\
BASIS(:,3)=C5 \ BASIS(:,2)\\
BASIS(:,4)=C5 \ BASIS(:,3)\\
BASIS(:,5)=C5 \ BASIS(:,4).
\end{array} \]
The five points of the plane corresponding to 
\[ BASIS(:,1),\quad BASIS(:,2),\quad BASIS(:,3),\quad BASIS(:,4),\quad BASIS(:,5) \]
are the vertices of a regular pentagon. If we add the points
\[ -BASIS(:,1),\quad -BASIS(:,2),\quad -BASIS(:,3),\quad -BASIS(:,4),\quad -BASIS(:,5)\]
we get the vertices of a regular decagon representing {\it the first shell of our
two-shell decagonal cluster} $\mathcal{C}$.\\
In a very similar way, by starting from the second given vector
\[ BASIS(:,6)=\left( \begin{array}{c}
BASIS(1,6)\\[2mm]
BASIS(2,6)
\end{array}\right)  \]
we generate the vectors
\[\begin{array}{l}
BASIS(:,7)\ =C5 \ BASIS(:,6) \\
BASIS(:,8)\ =C5 \ BASIS(:,7)\\
BASIS(:,9)\ =C5 \ BASIS(:,8)\\
BASIS(:,10)=C5 \ BASIS(:,9).
\end{array} \]
The 10 points
\[ \begin{array}{rrrrr}
BASIS(:,6),& BASIS(:,7),& BASIS(:,8),& BASIS(:,9),& BASIS(:,10) \\[2mm]
 -BASIS(:,6),& -BASIS(:,7),& -BASIS(:,8),& -BASIS(:,9),& -BASIS(:,10)
\end{array}\]
are the vertices of a regular decagon representing {\it the second shell of our
two-shell decagonal cluster} $\mathcal{C}$.\\[1cm]

{\large \bf 2. The physical two-dimensional space $\mathbb{E}$ regarded 
as a subspace of $\mathbb{R}^{10}$}\\[5mm]
The vectors of $\mathbb{R}^{10}$
\[ BASIS(1,:)=(\ BASIS(1,1), \, BASIS(1,2), \, BASIS(1,3), \ ... \ , BASIS(1,10)) \]
\[ BASIS(2,:)=(\ BASIS(2,1), \, BASIS(2,2), \, BASIS(2,3), \ ... \ , BASIS(2,10)) \]
are orthogonal
\[ \langle \ BASIS(1,:),\, BASIS(2,:) \ \rangle =
\sum_{J=1}^{10}\ BASIS(1,J)\ \cdot \ BASIS(2,J) \ =\ 0\]
and have the same norm
\[ NORM = \sqrt{\sum_{J=1}^{10}\ [\, BASIS(1,J)\, ]^2 }
         = \sqrt{\sum_{J=1}^{10}\ [\, BASIS(2,J)\, ]^2 }.\]
The subspace $\mathbb{E}$ of $\mathbb{R}^{10}$ spanned by these vectors is a 
two-dimensional space representing {\it the physical space}. The orthogonal basis
\[ \mathcal{B}=\{ \ BASIS(1,:),\ BASIS(2,:) \ \} \]
is very useful when we have to find the orthogonal projection 
$\mathcal{P}_\mathbb{E}V$ of a vector
$V\in \mathbb{R}^{10}$ on $\mathbb{E}$.\\ 
Since 
\[ \mathbb{B}=\left\{ \ \frac{BASIS(1,:)}{NORM},\  \frac{BASIS(2,:)}{NORM}\ \right\} \]
is an orthonormal basis of $\mathbb{E}$ we get  
\[ \mathcal{P}_\mathbb{E}V = 
\left\langle \ V,\ \frac{BASIS(1,:)}{NORM}\ \right\rangle \frac{BASIS(1,:)}{NORM}\ +
\left\langle \ V,\ \frac{BASIS(2,:)}{NORM}\ \right\rangle \frac{BASIS(2,:)}{NORM}\]
that is, the coordinates of $\mathcal{P}_\mathbb{E}V $ in the orthonormal basis
$\mathbb{B}$ are 
\[ \left( \  \left\langle \ V,\ \frac{BASIS(1,:)}{NORM}\ \right\rangle , \ 
       \left\langle \ V,\ \frac{BASIS(2,:)}{NORM}\ \right\rangle \ \right) \]
\[ =\frac{1}{NORM}\left( \ \langle \ V,\ BASIS(1,:) \ \rangle , \ 
       \langle \ V,\ BASIS(2,:)\ \rangle \ \right). \]
In our program we have to find the projection of the points of $\mathbb{Z}^{10}$
lying in the strip on $\mathbb{E}$.\\ In this case the factor $1/NORM$ is the same for
all the points (scaling factor) and we neglect it, that is, we use the projector
\[ \mathcal{P}:\mathbb{R}^{10}\longrightarrow \mathbb{R}^2\qquad
\mathcal{P}V=\left( \ \langle \ V,\ BASIS(1,:) \ \rangle , \ 
       \langle \ V,\ BASIS(2,:)\ \rangle \ \right). \]

{\large \bf 4. The window $\mathbb{W}$}\\[5mm]
Let \ $\mathcal{P}_\mathbb{E}^\perp : \mathbb{R}^{10}\longrightarrow \mathbb{E}^\perp $
be the orthogonal projector corresponding to the subspace
\[ \mathbb{E}^\perp =\{ \ V \in \mathbb{R}^{10}\ |\ \ 
\langle V,W \rangle =0\ {\rm for\ all}\ W\in \mathbb{E}\ \} \]
This 8-dimensional subspace (called the {\it internal space}) is the orthogonal
complement of $\mathbb{E}$. \\
The {\it window} $\mathbb{W}$ is the projection on $\mathbb{E}^\perp $
\[ \mathbb{W}=\mathcal{P}_\mathbb{E}^\perp \left( \ \mathbb{K}\ \right) \]
of the hypercube $\mathbb{K}=[ \ -0.5,\ 0.5\ ]^{10}$, that is,  
\[ \mathbb{K}=\{ \ (\ V(1),\, V(2), \, ...\, ,\, V(10)\ )\ |\ \ 
-0.5\leq V(J)\leq 0.5\ {\rm for\ all\ } J\in \{ 1,2,...,10\} \ \}. \]
The window $\mathbb{W}$ is a polyhedron in the 8-dimensional space $\mathbb{E}^\perp $.
The 7-dimensional faces of $\mathbb{W}$ are the projections of certain 7-dimensional 
faces of the hypercube $\mathbb{K}$ from $\mathbb{R}^{10}$.

Let 
\[ E(I,:)=(\, E(I,1),\, E(I,2),\, E(I,3),\, ...,\, E(I,10)\, )\qquad  
I\in \{ 1,2,3,...,10\} \]
be the vectors of the canonical basis of $\mathbb{R}^{10}$, that is,
\[ \begin{array}{l}
E(1,:)\ =(1,0,0,0,0,0,0,0,0,0)\\
E(2,:)\ =(0,1,0,0,0,0,0,0,0,0)\\
E(3,:)\ =(0,0,1,0,0,0,0,0,0,0)\\
......................................\\
E(10,:)=(0,0,0,0,0,0,0,0,0,1).
\end{array} \]
Each 7-face of $\mathbb{K}$ is parallel to 7 of these vectors 
and orthogonal to the other 3 vectors.\\ For each three distinct vectors
\[ E(I1,:),\quad E(I2,:),\quad E(I3,:)\] 
the number of 7-faces of $\mathbb{K}$ orthogonal to them is $2^3$.
The hypercube $\mathbb{K}$ has 
\[      \left( \begin{array}{c}
10\\
3\end{array}\right)=\frac{10\cdot 9\cdot 8}{1\cdot 2\cdot 3}=210\]
sets of $2^3$ parallel 7-faces. We label them by using the set
\[ \{ \ (I1,I2,I3)\in \mathbb{Z}^3\ \ |\ \ 1\leq I1 \leq 8, \ \ 
                                           I1+1\leq I2 \leq 9, \ \ 
                                           I2+1\leq I3 \leq 10\ \} \]
having 210 elements. The set
\[ \left\{ \ U \in \mathbb{R}^{10}\ \ \left| \ 
\begin{array}{l}
U(I1),\, U(I2),\, U(I3)\in \{ -0.5,\, 0.5\}\\
U(I)=0\ {\rm for}\ J\not\in \{ I1,\, I2,\, I3\}
\end{array} \right. \right\} \]
contains a point and only one from each of the eight 7-faces 
of $\mathbb{K}$ corresponding to $(I1,I2,I3)$.\\[1cm]

{\large \bf 3. The strip $\mathbb{S}$ and the quasiperiodic 
pattern $\mathcal{Q}$}\\[5mm]
The {\it strip} $\mathbb{S}$ is defined as
\[ \mathbb{S}=\{ \ V\in \mathbb{R}^{10}\ \ |\ \ \ 
\mathcal{P}_\mathbb{E}^\perp V\in \mathbb{W}\ \ \}. \]
In $\mathbb{R}^3$ the cross-product of two vectors 
$v=(v_x,v_y,v_z)$ and $w=(w_x.w_y,w_z)$ can be defined as
\[ v \times w = \left| \begin{array}{ccc}
\vec i & \vec j & \vec k \\
v_x & v_y & v_z\\
w_x & w_y & w_z
\end{array} \right|
=\left| \begin{array}{cc}
v_y & v_z\\
w_y & w_z
\end{array} \right| \vec i-
\left| \begin{array}{cc}
v_x & v_z\\
w_x & w_z
\end{array} \right| \vec j+
\left| \begin{array}{cc}
v_x & v_y\\
w_x & w_y
\end{array} \right| \vec k 
\]
where $\{ \vec i, \vec j, \vec k \}$ is the canonical orthonormal 
basis of $\mathbb{R}^3$.
The vector $v\times w$ is a vector orthogonal to $v$ and $w$, and 
the scalar product between this vector and any vector $u=(u_x,u_y,u_z)$
can be defined as
\[ \langle u,\, v \times w \rangle = \left| \begin{array}{ccc}
u_x & u_y & u_z\\
v_x & v_y & v_z\\
w_x & w_y & w_z
\end{array} \right|. \]
In a very similar way, we obtain a vector $W$ orthogonal to 9 vectors
\[ U(I,:)=(\, U(I,1),\,U(I,2),\, ...,\, U(I, 10)\, )\qquad
I\in \{ 1,2,3,...,9\} \]
by expanding the formal determinant
\[ W=\left|
\begin{array}{ccccc}
E(1,:) & E(2,:) & E(3,:) & ... & E(10,:)\\
U(1,1) & U(1,2) & U(1,3) & ... & U(1,10)\\ 
U(2,1) & U(2,2) & U(2,3) & ... & U(2,10)\\ 
... & ... & ... & ... & ...\\
U(9,1) & U(9,2) & U(9,3) & ... & U(9,10)
\end{array} \right| \]
containing in the first row the vectors of the canonical basis 
$\mathcal{B}$, and
\[ \langle V,W \rangle = \left|
\begin{array}{ccccc}
V(1) & V(2) & V(3) & ... & V(10)\\
U(1,1) & U(1,2) & U(1,3) & ... & U(1,10)\\ 
U(2,1) & U(2,2) & U(2,3) & ... & U(2,10)\\ 
... & ... & ... & ... & ...\\
U(9,1) & U(9,2) & U(9,3) & ... & U(9,10)
\end{array} \right| \]
for any vector $V$ from $\mathbb{R}^{10}$. For example,
\[ W=\left| \begin{array}{cccccccccc}
E(1,:) & E(2,:) & E(3,:) & E(4,:) & E(5,:) & 
E(6,:) & E(7,:) & E(8,:) & E(9,:) & E(10,:)\\
0 & 0 & 0 & 1 & 0 & 0 & 0 & 0 & 0 & 0\\ 
0 & 0 & 0 & 0 & 1 & 0 & 0 & 0 & 0 & 0\\ 
0 & 0 & 0 & 0 & 0 & 1 & 0 & 0 & 0 & 0\\ 
0 & 0 & 0 & 0 & 0 & 0 & 1 & 0 & 0 & 0\\ 
0 & 0 & 0 & 0 & 0 & 0 & 0 & 1 & 0 & 0\\ 
0 & 0 & 0 & 0 & 0 & 0 & 0 & 0 & 1 & 0\\ 
0 & 0 & 0 & 0 & 0 & 0 & 0 & 0 & 0 & 1\\
B(1,1) & B(1,2) & B(1,3) & B(1,4) & B(1,5) & 
B(1,6) & B(1,7) & B(1,8) & B(1,9) & B(1,10)\\
B(2,1) & B(2,2) & B(2,3) & B(2,4) & B(2,5) & 
B(2,6) & B(2,7) & B(2,8) & B(2,9) & B(2,10)
\end{array} \right| \]
\[ = - \left| \begin{array}{ccc}
E(1,:) & E(2,:) & E(3,:)\\
BASIS(1,1) & BASIS(1,2) & BASIS(1,3)\\
BASIS(2,1) & BASIS(2,2) & BASIS(2,3)
\end{array} \right| \]
where
\[ B(I,J)=BASIS(I,J) \]
is a vector orthogonal to the 9 vectors
\[ E(4,:),\  E(5,:),\  E(6,:),\  
E(7,:),\  E(8,:),\  E(9,:),\ E(10,:),\ BASIS(1,:)\ {\rm and}\  BASIS(2,:).\]
Since \ $E(I,:)-\mathcal{P}_\mathbb{E}^\perp E(I,:)$ is a linear 
combination of $BASIS(1,:)$ and $BASIS(2,:)$ it follows  that $W$ is a
vector lying in $\mathbb{E}^\perp $ orthogonal to the 7 vectors
\[ \mathcal{P}_\mathbb{E}^\perp E(4,:),\  
\mathcal{P}_\mathbb{E}^\perp E(5,:),\  
\mathcal{P}_\mathbb{E}^\perp E(6,:),\  
\mathcal{P}_\mathbb{E}^\perp E(7,:),\  
\mathcal{P}_\mathbb{E}^\perp E(8,:),\  
\mathcal{P}_\mathbb{E}^\perp E(9,:)\ {\rm and}\  
\mathcal{P}_\mathbb{E}^\perp E(10,:)\]
that is, ortogonal to the two parallel 7-faces of $\mathbb{W}$ corresponding to 
$(I1,I2,I3)=(1,2,3)$.\\
In addition, we get 
\[ \langle V,W\rangle =
\left| \begin{array}{ccc}
V(1) & V(2) & V(3)\\
BASIS(1,1) & BASIS(1,2) & BASIS(1,3)\\
BASIS(2,1) & BASIS(2,2) & BASIS(2,3)
\end{array} \right| \]
for any 
\[ V=(\, V(1),\, V(2),\, V(3),\, V(4),\, V(5),\, V(6),\, V(7),\, 
V(8),\, V(9),\, V(10)\, )\in \mathbb{R}^{10}\] 
The projection on $\mathbb{E}^\perp $ of a point $V\in \mathbb{Z}^{10}$
is ``between'' the two parallel 7-faces of $\mathbb{W}$ corresponding
to $(I1,I2,I3)=(1,2,3)$ if and only if
\[ 
-STRIP(1,2,3)\ \leq \ \left| \begin{array}{ccc}
V(1) & V(2) & V(3)\\
BASIS(1,1) & BASIS(1,2) & BASIS(1,3)\\
BASIS(2,1) & BASIS(2,2) & BASIS(2,3)
\end{array} \right| \ \leq \ STRIP(1,2,3) \]
where 
\[ STRIP(1,2,3)=\max_{ \footnotesize
\begin{array}{l}
U(1)\in \{ -0.5,\, 0.5\}\\
U(2)\in \{ -0.5,\, 0.5\}\\
U(3)\in \{ -0.5,\, 0.5\}
\end{array} }
\left| \begin{array}{ccc}
U(1) & U(2) & U(3)\\
BASIS(1,1) & BASIS(1,2) & BASIS(1,3)\\
BASIS(2,1) & BASIS(2,2) & BASIS(2,3)
\end{array} \right|.  \]
Similar relations can be obtained for each of the 120 systems of 
indices (I1,I2,I3).\\

The quasiperiodic pattern defined by the above construction is
\[ \mathcal{Q}=\{ \ \mathcal{P}V\ \ |\ \ 
V\in \mathbb{S}\cap \mathbb{Z}^{10}\ \} .\]
We have
\[ \mathcal{P}E(J,:)=\left( \ \langle \ E(J,:),\ BASIS(1,:) \ \rangle , \ 
       \langle \ E(J,:),\ BASIS(2,:)\ \rangle \ \right)=
(\, BASIS(1,J),\, BASIS(2,J)\, ) .\]
Therefore, for each point $\mathcal{P}V$ of $\mathcal{Q}$ the nearest 
and second nearest neighbours are distributed at (some of) the
vertices of the translation $\mathcal{P}V +\mathcal{C}$
of the generating two-shell cluster $\mathcal{C}$.
The pattern $\mathcal{Q}$ can be regarded as a quasiperiodic packing of 
copies of the cluster $\mathcal{C}$.\\

More details, bibliography and samples can be found on the website
\begin{center}
http:\verb#//fpcm5.fizica.unibuc.ro/~ncotfas#
\end{center}

{\large \bf 5. Computer program in FORTRAN and MATHEMATICA}\\
\begin{verbatim}
! QUASIPERIODIC PACKING OF COPIES OF A TWO-SHELL DECAGONAL CLUSTER
! ****************************************************************

! PLEASE INDICATE HOW MANY POINTS DO YOU WANT TO ANALYSE 
      INTEGER, PARAMETER :: N = 10000 

	  INTEGER I, J, K, L, I1, I2, I3, JJ, JP, JPP 
	  REAL D1, D2, D3, R, XP, YP						  
      REAL, DIMENSION(10) :: V, W, TRANSLATION			  
	  REAL, DIMENSION(2,10) :: BASIS			
	  REAL, DIMENSION(2,2) :: C5 
	  REAL, DIMENSION(1:8,2:9,3:10) :: STRIP
	  REAL, DIMENSION(N ,10) :: POINTS
	  REAL, DIMENSION(N) :: XPOINT, YPOINT
	    
! PLEASE INDICATE THE COORDINATES OF A POINT BELONGING TO THE FIRST SHELL
      BASIS(1,1) = 1.0   
	  BASIS(2,1) = 0.0  

! PLEASE INDICATE THE COORDINATES OF A POINT BELONGING TO THE SECOND SHELL
      BASIS(1,6) = 0.9   
	  BASIS(2,6) = 1.1  

! PLEASE INDICATE THE TRANSLATION OF THE STRIP YOU WANT TO USE	  
	  TRANSLATION = 3.7 
	                      
	  C5(1,1) = (SQRT(5.0)-1.0)/4.0	 
	  C5(1,2) = -SQRT(5.0 + SQRT(5.0))/(2.0 * SQRT(2.0))
	  C5(2,1) =  SQRT(5.0 + SQRT(5.0))/(2.0 * SQRT(2.0))
	  C5(2,2) = (SQRT(5.0)-1.0)/4.0
	  DO J = 2, 5
	   DO I = 1, 2
        BASIS(I,J) = C5(I,1) * BASIS(1,J-1) + C5(I,2) * BASIS(2,J-1)
	    BASIS(I,5+J) = C5(I,1) * BASIS(1,4+J) + C5(I,2) * BASIS(2,4+J)
       END DO	 
      END DO	
      STRIP=0 
      DO I1 =1, 8		
	  DO I2 =I1+1, 9	  
	  DO I3 =I2+1, 10	  
	    DO D1 =-0.5, 0.5  
	    DO D2 =-0.5, 0.5  
	    DO D3 =-0.5, 0.5
	    R = D1 * BASIS(1,I2) * BASIS(2,I3) + D3 * BASIS(1,I1) * BASIS(2,I2) + &
		    D2 * BASIS(1,I3) * BASIS(2,I1) - D3 * BASIS(1,I2) * BASIS(2,I1) - &
		    D1 * BASIS(1,I3) * BASIS(2,I2) - D2 * BASIS(1,I1) * BASIS(2,I3)
	    IF ( R > STRIP(I1,I2,I3) ) 	STRIP(I1,I2,I3) = R
	    END DO
	    END DO	  
		END DO	 
		IF( STRIP(I1,I2,I3) .EQ. 0 ) STRIP(I1,I2,I3)=N * SUM( BASIS(1,:) ** 2)
	  END DO	  
	  END DO	 
	  END DO	 
	  PRINT*, 'COORDINATES OF THE POINTS OF THE TWO-SHELL C5-CLUSTER:'
	  DO J = 1, 10
	  PRINT*, J, BASIS(1,J), BASIS(2,J)
	  END DO
	  PRINT*, '* STRIP TRANSLATED BY THE VECTOR WITH COORDINATES:'
	  PRINT*,    TRANSLATION
	  PRINT*, '* PLEASE WAIT A FEW MINUTES OR MORE, & 
                   DEPENDING ON THE NUMBER OF ANALYSED POINTS'
	  JP = 0        
   	  POINTS = 0
	  POINTS(1,:) = ANINT( TRANSLATION)
	  K = 1			        
	  L = 0				
      DO I = 1, N
   	  V = POINTS(I, : )	- TRANSLATION
      JJ = 1				
	  JPP = 0
      DO I1 =1, 8			
	   DO I2 =I1+1, 9		
	    DO I3 =I2+1, 10
	R = V(I1) * BASIS(1,I2) * BASIS(2,I3) + V(I3) * BASIS(1,I1) * BASIS(2,I2) + &
	    V(I2) * BASIS(1,I3) * BASIS(2,I1) - V(I3) * BASIS(1,I2) * BASIS(2,I1) - &
		  V(I1) * BASIS(1,I3) * BASIS(2,I2) - V(I2) * BASIS(1,I1) * BASIS(2,I3)
          IF ( R < - STRIP(I1,I2,I3) .OR. R > STRIP(I1,I2,I3)) JJ = 0
		  IF ( R == - STRIP(I1,I2,I3) .OR. R == STRIP(I1,I2,I3)) JPP = 1
		 END DO			
	   END DO			
	  END DO	 	
      IF( JJ .EQ. 1 ) THEN
	  XP = 	SUM( V * BASIS(1,:) )	
	  YP = 	SUM( V * BASIS(2,:) )
	  I3 = 1
	  DO J = 1, L
	  IF( XP == XPOINT(J) .AND. YP == YPOINT(J) ) I3 = 0
	  END DO
	  IF( I3 == 1) THEN	
	  IF( JPP .EQ. 1 ) JP = JP + 1	 
      L = L + 1							  
      XPOINT(L) = XP	  
      YPOINT(L) = YP
	  ELSE
	  END IF  
      DO I1 = 1, 10			  
	   DO I2 = -1, 1		  
	     W = POINTS(I, : )				 
	     W(I1) = W(I1) + I2
	     I3 = 0									
	      DO J = 1, K							 
          IF( ALL(W .EQ. POINTS(J,:)) ) I3 = 1	 
	      END DO								 
	       IF ( I3 == 0 .AND. K < N ) THEN	  
	       K = K + 1					
	       POINTS(K, : ) = W				
	       ELSE							
	       END IF
	    END DO						   
	   END DO						   
	   ELSE							 
	   END IF						 
      END DO
	  PRINT*, 'NUMBER OF ANALYSED POINTS :', K
      PRINT*, 'NUMBER OF OBTAINED POINTS :', L
	  PRINT*, 'NUMBER OF POINTS LYING ON THE FRONTIER OF THE STRIP:', JP
	  PRINT*, 'PLEASE INDICATE THE NAME OF A FILE FOR RESULTS'
	  WRITE(4,10)					  
  10  FORMAT('Show[Graphics[{PointSize[0.03], { ') 
	  DO J = 1, 10
	  WRITE(4,20) BASIS(1,J), BASIS(2,J)		
  20  FORMAT( 'Point[{'F10.5','F10.5,'}], ')
	  END DO
	  DO J = 1, 10
	  WRITE(4,30) -BASIS(1,J), -BASIS(2,J)		
  30  FORMAT( 'Point[{'F10.5','F10.5,'}], ')
	  END DO
	  WRITE(4,40) 0.0, 0.0
  40  FORMAT( 'Point[{'F10.5','F10.5,'}]} }], &
                             PlotRange -> All, AspectRatio -> 1 ]')
	  WRITE(4,80)					  
  80  FORMAT('Show[Graphics[{PointSize[0.02], { ')
	  DO J = 1, L-1
	  WRITE(4,90) XPOINT(J), YPOINT(J)		
  90  FORMAT( 'Point[{'F10.5','F10.5,'}], ')
	  END DO
  	  WRITE(4,100) XPOINT(L), YPOINT(L)
 100  FORMAT( 'Point[{'F10.5','F10.5,'}]} } ], &
                               PlotRange -> All, AspectRatio -> 1]')
      PRINT*, '* OPEN THE FILE CONTAINING THE RESULTS WITH "NotePad" '
	PRINT*, '* SELECT THE CONTENT OF THE FILE ("Select All") & 
                                             AND COPY IT ("Copy")'
	PRINT*, '* OPEN "MATHEMATICA", PASTE THE COPIED FILE, &
                                       AND EXECUTE IT ("Shift+Enter").' 
	END
\end{verbatim}  
\end{document}